\input phyzzx
\centerline{GRAVITATION, THE QUANTUM, AND COSMOLOGICAL CONSTANT}
\vskip .5cm
\centerline{Pawel O. Mazur}
\vskip .1cm
\centerline{Department of Physics and Astronomy, }
\vskip .1cm
\centerline{University of South Carolina, Columbia, SC 29208, USA}
\foot{E-mail address: Mazur@psc.psc.sc.edu}
\foot{Submitted to Acta Physica Polonica {\bf B} in January 1996.}
\foot{Contribution to 
the special issue of Acta Physica Polonica {\bf B} celebrating 
Professor Andrzej Bialas' 60th birthday.}\foot{The expanded version of this 
work is about to appear in another journal. This material was also 
presented in two talks at UCLA and CALTECH on February 22-23, 1996.}
\vskip .5cm
\centerline{Abstract}

 The arguments of statistical nature for the existence 
of constituents of active gravitational masses are presented. 
The present paper proposes a basis for microscopic theory of universal 
gravitation. Questions like the relation of cosmological constant and 
quantum theory, black hole radiance and the nature of inertia are addressed. 
This paper is the second in the series of papers published in Acta Physica 
Polonica {\bf B}.
\vfill

  It is well known that the classical gravitating systems behave in the 
way foreign to statistical quantum mechanics. The negative specific heat of 
those systems and the phenomenon of gravitational collapse are different 
facets of the same reality. The difficulties arise which necessitate that 
the complete atomic description of gravitation and space-time be sought 
after. The departure from Einstein's General Relativity Theory compatible 
with the statistical basis of the Universal Second Law of Thermodynamics 
takes the form of the mass-energy quantization condition. 
In this paper we show that a gravitating mass $M$ in thermal equilibrium, 
sometimes called a black hole, behaves statistically like a system of 
some number $N$ of harmonic oscillators [1,2]  whose 
zero-point energy ${\epsilon\over 2}$ depends on $N$ universally 
in such a way that 

$$N{\epsilon}^2\sim {{hc^5}\over G}       , \eqno(1)$$

where $G$ is the Newton constant, $c$ is the velocity of light 
in vacuum, and $h$ is the Planck constant. 
The sum over all oscillators of the squares of zero-point energies is fixed 
and independent of the number of those oscillators. Similarly, the sum over 
the fourth power of zero-point energies gives the vacuum energy 
density, the so-called cosmological constant $\lambda$. 
The cosmological constant is identified by the order of 
magnitude with the number of oscillators $N$ in the Universe 

$$\lambda\sim N{\epsilon}^4(hc)^{-3}\sim {c^7\over G^2hN}    .   \eqno(2)$$

The lower bound for this total number of oscillators in the observed Universe 
is obtained from the present upper bound on $\lambda$, $N\sim 10^{122}$.   

 The system of $N$ correlated oscillators [2,3,15], known macroscopically as 
a gravitating mass, at the temperature of the cosmic relic radiation 
$T\sim 2.7 K^{\circ}$ has a total mass-energy exceeding the critical 
mass-energy of the observed Universe by $\sim 30$ orders of magnitude. 
This suggests that the observed Universe is extremely `young' and its age 
is only $10^{-15}$ times that of the time(-space) scale available to 
the whole Universe out there. Consider a system of $N$ 
{\it gravitational atoms} or {\it parcels of mass-energy} [2,3,12,15]. 
When $N$ is very large we must apply statistical reasoning to describe the 
state of such a large system. 
The statistical quantum theory is a statement about the objective reality 
based on the {\it Atomic Hypothesis} [2,3,12,15]. 
The statistical reasoning applied to the {\it simplest} properties of 
{\it gravitational atoms} leads to thermal properties of the mass-energy, 
which in the limit $N\rightarrow\infty$ converge to those already proposed 
earlier on the basis of the General Relativity Theory [4] 
and the Quantum Theory [5]. The enigmatic Bekenstein entropy 
of black holes [4] has not yet been derived 
on the basis of microscopic theory. Our work should be considered as 
a first step in the direction of establishing such a basis. The problem with 
previous approaches has been the silent assumption that the total entropy of 
black holes {\it must} be given by the Bekenstein formula [4], 

$$S_{bh} = 4k{\pi}M^2   ,    \eqno(3)$$

where $S_{bh}$ is the entropy, $k$ is the Boltzmann constant, and $M$ 
is the mass-energy in the Planck units. The postulate of gravitational 
constituents (oscillators, quanta, etc.) leads to Bekenstein's formula [4]  
only after a part of mass-energy fluctuations is neglected. 
We will show that neglecting ${1\over N}$ terms 
in the mass-energy fluctuation formula of Einstein [1], 
when applied to the system consisting of $N$ {\it gravitational atoms}, 
leads to the thermodynamics of black holes [4,5,3]. 

 The principal conclusion from the {\it Atomic Hypothesis} 
is that the thermodynamical limit of infinitely large $N$ at fixed mass-energy 
eliminates the additional source of mass-energy fluctuations 
which is responsible for the total positive heat capacity of gravitating 
masses. Neglecting this new kind of mass-energy fluctuations leads to one 
aspect of phenomena known macroscopically as the universal gravitation. 
This is much like neglecting the famous first term in the energy fluctuation 
formula for the 
black body radiation which was discovered by Einstein [1]. The true 
meaning of the first term, as we know today, is the 
corpuscular nature of electromagnetic phenomena. The second term in the 
fluctuation formula for black body radiation is the well known 
Rayleigh-Jeans term. 
Therefore, neglecting the first term in the energy fluctuation 
formula for the black body radiation amounts to an incomplete 
description of radiation as waves. 
{\it The direct implication of neglecting a} ${1\over N}$ 
{\it term in the formula for the gravitational mass-energy fluctuations 
is that the specific heat is negative for gravitating systems.} 
This also means that {\it the gravitational collapse,} 
with its typical property of negative specific heat 
(the latter one is governing formation of stars, galaxies, clusters 
and superclusters of galaxies etc.), 
{\it can proceed only to the extent that the 
new source of mass-energy fluctuations is neglected completely.} 
If, however, the circumstances occur that those 
new mass-energy fluctuations cannot be neglected, then a completely new 
physical reality opens up. The collapse is avoided. We should expect 
a plethora of new phenomena where those new mass-energy fluctuations 
shall demonstrate themselves. 

 In this paper we shall describe some predictions about phenomena 
following from physical reasoning based on {\it Atomic Hypothesis}. 
One of them concerns the cosmological constant [6]. 
The paradox of large cosmological constant was noticed pretty late [7,8]. 
It appears to be a paradox because the present theoretical models seem to have 
a limited scope of application. There is no doubt that 
the nonvanishing cosmological constant $\lambda$ must be very small. 
The present upper bound on the value of this constant is 
an extremely small number in natural units. We now know [9] that 
$\lambda$ cannot be larger than $10^{-122} \mu^4$, 
$\mu = \sqrt{hc\over G} = 5.46 10^{-5}g$. 

 The smallness of the cosmological constant 
in natural Planck units is a result of an almost 
perfect thermodynamical limit. This is to say that the smallness of the 
cosmological constant is an effect due to an enormous number $N$ 
of hypothetical {\it gravitational atoms}. 
The present upper bound on the `cosmological constant' $\lambda$ allows us 
to draw the conclusion about the lower bound on a number of 
{\it gravitational atoms} in the observed Universe, $N\sim 10^{122}$. 
The large numbers' coincidences noticed long time ago [10,11] 
have something in it after all. 
We have only one reservation to add: 
elementary particles are not the same as the hypothetical 
{\it gravitational atoms}, and, therefore, 
have little to do with the cosmological constant. 
The existence of the latter must be inferred indirectly from phenomena as it 
was previously done for atoms and elementary particles. 
We find an unusual coordination between the gravitational atomistic 
aspect of physical reality in the regime usually called infrared, or large 
distance scale, and the regime usually called ultraviolet, or short distance 
scale [17]. 

 It is now time to turn to the elementary physical considerations which have 
led the present author to the general {\it Atomic Hypothesis} [2,3,12,15,18]. 
Consider $N$ identical harmonic oscillators, characterized by 
the zero-point energy ${1 \over 2}\epsilon$. 
We assume that $\epsilon$ depends on $N$ in such a way 
that when $N\rightarrow\infty$ then $\epsilon(N)\rightarrow 0$. 
This assumption is logically simple, and also necessary because otherwise 
the generalized correspondence principle is violated [2,3]. 
The thermal properties of a large gravitating mass would be in disagreement 
with the predictions following from General Relativity Theory 
and Quantum Mechanics [4,5] unless the $N = \infty$ limit is taken 
in the formulas for the specific heat capacity with a fixed mass-energy. 
We assume that for large $N$ the 
following formula is valid: 

$$N{\epsilon}^2 = b\mu^2c^4   ,      \eqno(4)$$

where $\mu = \sqrt{hc \over G}$ is the Planck mass-energy. 
$b$ is a numerical constant 
to be determined later from the correspondence principle considerations. 
This formula ought to be justified by the results and predictions 
following from it, in the same way the hypothesis of 
the universal law of gravitation was justified by the derivation 
of the Kepler laws. The mathematical theory of new wave equations 
for gravitating particles [2,3,12,15,18] 
leads to mass-energy quantization of the type assumed in this paper. 
 
 Consider now the partition function of $N$ correlated harmonic oscillators 

$$Z(N,\beta) = \biggl(2sinh{\beta \epsilon'(N) \over 2}\biggr)^{-N}   ,  
\eqno(5)$$

where $\epsilon' = \epsilon - \chi$, and $\chi$ is the chemical potential. 

In the following we assume ${\epsilon}'>0$. 
The average mass-energy of this system is 

$$\overline{E} = U = N{\epsilon}\biggl({1 \over 2} + {1 \over 
{e^{\beta\epsilon'} - 1}}\biggr)  ,         \eqno(6)$$

$\beta = {1\over kT}$. 

Now, the Einstein energy fluctuation formula [1], 

$$\overline{(\Delta E)^2} = -{\partial U \over \partial \beta}  ,   \eqno(7)$$ 

gives for our system the following expression 

$$\overline{(\Delta E)^2} = N\epsilon^2 \biggl(
{1 \over {e^{\beta\epsilon'} - 1}} + 
{1\over (e^{\beta\epsilon'} - 1)^2}\biggr)  .          \eqno(8)$$
 
The average mass-energy at zero temperature is 

$$U_0 = {N\epsilon \over 2}   .         \eqno(9)$$

Defining the deviation of an average mass-energy $U$ at a given 
temperature from its zero temperature value $U_0$ by 

$$\Delta U = U - U_0 = 2U_0(e^{\beta\epsilon'} - 1)^{-1}   ,   \eqno(10)$$
 
we see that the energy fluctuations are of two 
types: there are terms linear and quadratic in $\Delta U$. The linear term 
is corresponding to the corpuscular character of quanta (it is of 
an order of $N^{-{1\over 2}}$), while the quadratic 
term corresponds to their wave character (it is of the order of $N^{-1}$). 
This formula has been well known for the last $90$ years [1]. 
The gravitational mass-energy displays not only the 
corpuscular characteristic, as described by the General Theory of Relativity, 
but also the wave-like behaviour typical of wave mechanics. 
The conclusions we can draw from the first principles are twofold. 
The fact that the gravitational mass-energy 
coordinates all three fundamental categories of existence of an objective 
reality, time, space, and matter, leads us to suggest the following: 
In addition to {\it `geometric optics'}, or corpuscular, 
behaviour predicted by the General Theory of Relativity there exists 
a new {\it `wave optics'} regime of behaviour of space-time-matter. 
{\it Not only the mass-energy displays the wave-like 
property but also the space and time seem to behave in this way, as far as the 
statistical properties of gravitating systems are concerned.} 
How this fact coordinates with the observed properties of gravitating masses? 

 The most unusual character of the gravitational mass-energy 
oscillators is that they somehow manage, via the quadratic 
sum rule defining the Newton constant $G$, 

$$\sum_i {\epsilon_i}^2 = b{hc^5 \over G}  ,              \eqno(11)$$

to reduce their zero-point energy when the number $N$ of them grows. 
This also means that a cold large gravitational mass $M\sim \mu{\sqrt N}$ 
consists of $N$ {\it constituents}. The formula 

$$M^2 = {1\over {2\pi}}\mu^2N  ,  \eqno(12)$$ 

was derived long time ago by the present author [12]. 
The physical meaning of the `phenomenological' entropy of Bekenstein [4] 
is that it is the measure of the number $N$ 
of {\it constituents} making up a very cold large body. The more 
massive is a gravitating mass the softer are the {\it constituents} 
or {\it gravitational quanta}. Otherwise, as usual with oscillators, there are 
two sources of statistical fluctuations of mass-energy corresponding to the 
corpuscular and wave aspect of quanta. We calculate the mass-energy 
fluctuations of the system in terms of the average energy $U$, 

$$\overline{(\Delta E)^2} = N^{-1}U^2 - {1 \over 4}N{\epsilon}^2 = 
N^{-1}U^2 - {1 \over 4}b{\mu}^2c^4   .          \eqno(13)$$ 

Neglecting the ${1 \over N}$ term in this formula, when $U$ is fixed, 
we obtain the expression for statistical fluctuations typical of gravitating 
systems: 

$$\overline{{(\Delta E)^2}_{bh}} = -{1\over 4}b{\mu}^2c^4  .      \eqno(14)$$ 

The well known relation between the mass-energy fluctuations 
and the behaviour of entropy near the state of thermal equilibrium, 

$$\overline{(\Delta E)^2} = 
- k\biggl({\partial^2 S \over \partial U^2}\biggr)^{-1}   ,      \eqno(15)$$ 

leads to the entropy of such a {\it truncated system}: 

$${\partial^2 S_{bh} \over \partial U^2} = 4kb^{-1}{\mu}^{-2}c^{-4}  .     
\eqno(16)$$ 

Integrating this last equation gives the inverse temperature 

$$\beta_{bh} = 4b^{-1}\mu^{-2}c^{-4}U  ,            \eqno(17)$$ 
where an arbitrary integration constant 
is fixed to be zero by demanding that a very massive body is also 
very cold [4,5]. The entropy is given by the 
`phenomenological' entropy formula of Bekenstein [4]: 

$$S_{bh} = 2kb^{-1}{\mu}^{-2}c^{-4}U^2 + const  .       \eqno(18)$$

The model calculation of Hawking [5] leads to a numerical 
value of the constant $b$, $b = {1 \over 4{\pi}^2}$. 
Quite independently of the 
actual value of the numerical constant $b$ the entropy $S_{bh}$ has a lower 
bound 

$$S_0 = 2kb^{-1}{\mu}^{-2}c^{-4}{U_0}^2   ,        \eqno(19)$$ 

which depends only on $N$. This follows from the fact that 
the total mass-energy $U$ is bounded from below by the zero temperature 
value $U_0$, $U\geq U_0$. Now, 

$${U_0}^2 = {1\over 4}b\mu^2c^4N   ,             \eqno(20)$$

and, therefore, the lower bound on the entropy does not depend on $b$ at all, 

$$S_0 = k{N \over 2}  .                    \eqno(21)$$ 

It is quite natural for the entropy to be bounded 
from below by the number ${N \over 2}$ of {\it constituents}.  

 We have seen the emergence of the Bekenstein formula [4] for the 
black hole entropy from the hypothesis about the microscopic nature of 
gravitational phenomena. It should be noticed 
that exactly in the same way as the light quanta [1] 
have emerged from the Wien black body radiation formula, 
the necessity of introduction of the {\it gravitational mass-energy quanta} 
is forced upon us by the Universal Second Law of Thermodynamics 
and the mass-energy fluctuation formula following from it. 
The adiabatic invariance arguments for the irreducible mass $M_{ir}$ of 
the Kerr black hole due to Christodoulou [13] 
have led to the concept of the black hole entropy of Bekenstein [4]. 
Bekenstein has proposed the Generalized Second Law of Thermodynamics [4], 
which states that the total entropy of a black hole and its exterior 
cannot decrease. 
The Universal Second Law of Thermodynamics [15] 
allows one to draw conclusions about the behaviour 
of a general system near its state of thermal equilibrium. 
In particular, the total entropy {\it must be} 
a {\it maximum} at the state of thermal equilibrium. 
The Boltzmann formula, 

$$S = klnW   ,                \eqno(22)$$

coordinating the relation between 
the thermodynamical property of a system, 
the entropy $S$, and the {\it thermodynamic probability} $W$ allows us 
to draw conclusions about the total combinatorial factors defining $W$ 
in terms of statistics of {\it atoms} or {\it quanta}. 
The positivity of $\overline{(\Delta E)^2}$ is strictly implied 
by the maximal value of the Boltzmann thermodynamic 
probability $W$ at the state of thermal equilibrium. 
We have applied this idea to {\it gravitational atoms}. 
The thermodynamic probability $W$ calculated on the basis of 
our hypothesis poses the following question: 
What kind of statistics leads to this $W$? I will report on this question 
later. 

 If the Bekenstein entropy were the whole thing, as far as the thermal 
properties of gravitating masses are concerned, then 
the World would be always in a state of the lowest thermodynamic 
probability. This conclusion would lead then to the statement that 
the behaviour of a visible Universe is determined by the condition 
that it is in a state of the lowest statistical weight. 
Considering an ensemble of such Universes, regarded as local 
thermal phenomena in a sense suggested in the introduction, we would be 
persuaded to conclude that our Universe is the least probable one. 
The Universe must be regarded as a very typical one in the 
{\it statistical ensemble of Universes}, which is also the statement 
of the maximal thermodynamic probability $W$ of Boltzmann.  
The Universal Second Law of Thermodynamics 
with its property of positive mass-energy fluctuations which we have 
consistently used in our arguments for the {\it Atomic Hypothesis} 
[2,3,12,15] must be considered as the basic notion underlying 
the Law of Universal Gravitation. It should be noticed 
that the notion of a statistical ensemble for the observable Universe 
is justified only after we identify {\it atoms} 
whose existence is underlying the totality of phenomena. 

 We have given an independent statistical arguments for the existence 
of {\it gravitational atoms} elsewhere [3,15]. 
The arguments advanced, when taken at the face value, 
mean that the observed negative specific heat of gravitating systems 
is a result of a coarse grained description of the phenomena. 
Apparently there exist statistical fluctuations in the mass-energy 
whose role is to compensate in some regime the 
negative contribution coming from the large scale part of the fluctuation 
spectrum, with the latter observed on a macroscopic scale. 
The fact that the other, compensating, part of the spectrum of fluctuations 
is not observed at large scale does not mean that it is not inherent in 
phenomena when inspected closely. 

 One of the most obvious implications of the atomistic nature of gravitation 
is that large massive objects believed to be formed in the 
gravitational collapse will not display the total negative specific heat 
property.{\it In fact, in contrary to the prediction about the nature of an 
almost thermal radiation emitted by black holes} [5], 
{\it those objects do not get hotter in the process of losing energy.} 
Quite opposite behaviour takes place: 

{\it objects initially very hot become cooler and cooler as they emit energy.}

In particular,{\it we should not expect to observe 
mini-black hole explosions} [5,14] {\it at all}. 
{\it In fact, there exist at least two ways 
to understand the lack of observational evidence for primordial 
black hole explosions. One of them is that such black holes were 
not produced copiously in a very early hot periods of existence of our 
Universe, which is also a hypothesis of small probability. The second one 
is that very dense objects with some spectrum of masses were produced 
copiously in an early Universe} [16,14] {\it but they became cooler by losing 
enormous amounts of energy to the surrounding space. It is, therefore, 
not unlikely that the most distant quasars may give us information on the 
dynamics of energy production and its emission mechanisms which will be 
compatible with the hypothesis presented here. }
{\it The younger the quasars the hotter they should appear.} 
This means that {\it for the highest redshift} $\Delta z$ 
{\it quasars we should expect on average 
the highest power of emission of energy (the highest total luminosity).} 
Obviously we need more detailed models built on the 
basis of {\it Atomic Hypothesis} about the nature of gravitation in order 
to be able to give more quantitative predictions. 

 We came to the realisation that a gravitational mass-energy consists 
of a number of {\it constituents} [2,3,12,15]. 
This situation we find analogous to that one of a container of gas. 
Now, a gas container is considered isolated if the total number 
of gas molecules is sustained constant over 
the period of time. As far as the {\it mass-energy parcels} are concerned 
a `container' consisting of a given number $N$ of {\it gravitational atoms} 
has a total mass-energy $E\sim \mu\sqrt{N}$ (at zero temperature). 

A `container' of {\it gravitational atoms} 
is considered isolated if the number of atoms $N$, and, therefore, 
its mass-energy $E$ is constant. The change in $N$ is the measure of motion. 
Now, we can formulate the First Newton's Law 
of Inertia in the following way: 

 {\it The physical system called an inertial mass-energy exists in a state 
characterized by a constant number of gravitational atoms. 
When this number of atoms is changing 
the state of motion is changing. Under such circumstances we say that 
there are forces acting on an inertial mass-energy.} 

 We need the proper formal language of new difference wave equations 
[2,3,15] for gravitating particles which would allow us a more detailed 
knowledge of microscopic processes underlying gravitation, and, therefore, 
space-time. We hope to report on the progress in this direction in due time. 

\centerline{Aknowledgement}

 I would like to thank Professors A. Staruszkiewicz, A. Z. G\'orski, 
J. Szmigielski, Y. Aharonov, S. Nussinov, J. Cornwall, E. Mottola, 
I. Antoniadis, K. Narain and J. Schwarz for discussions concerning 
the content of this paper. 

 This research was partially supported by NSF grant to University of South 
Carolina. 
\vskip 2cm
\centerline{REFERENCES}

\item{ 1}   A. Einstein, Ann. Phys. {\bf 17}, 132 (1905); 
ibid. {\bf 20}, 199 (1906); ibid. {\bf 22}, 180 (1907); 
Physik. Zeit. {\bf 10}, 185, 817 (1909).
\item{ 2}   P. O. Mazur, Acta Phys. Polon. {\bf B26}, 1685-1697 (1995).
\item{ 3}   P. O. Mazur, submitted to Physical Review Letters, February 1996.
\item{ 4}   J. D. Bekenstein, Phys. Rev. {\bf D7}, 2333-2346 (1973).
\item{ 5}   S. W. Hawking, Nature {\bf 248}, 30-31 (1974); 
Commun. Math. Phys. {\bf 43}, 199-222 (1975).
\item{ 6}   A. Einstein, S.-B. Preuss. Akad. Wiss. {\bf 142} (4) (1917). 
\item{ 7}   Ya. Zel'dovich, Pisma v ZETF {\bf 6}, 883-884 (1967); 
see also vol. 2 of {\it Selected Works} p.165, Princeton University Press, 1993.
\item{ 8}   A. D. Sakharov, Dokl. Akad. Nauk SSSR {\bf 177}, 70-71 (1967); 
Sov.Phys. Dokl. {\bf 12}, 1040-1041 (1968). 
\item{ 9}   S. Weinberg, Rev. Mod. Phys. {\bf 61}, 1-51 (1989).
\item{10}   P. A. M. Dirac, Proc. Roy. Soc. London {\bf A165}, 199-208 (1938).
\item{11}   A. S. Eddington, Proc. Roy. Soc. London {\bf 133}, 605-614 (1931).
\item{12}   P. O. Mazur, GRG{\bf 19}, 1173-1180 (1987).
\item{13}   D. Christodoulou, Phys. Rev. Lett. {\bf 25}, 1596-1597 (1970).
\item{14}   B. Carr, Astrophys. Jour., {\bf 201}, 1-19 (1975). 
\item{15}   P. O. Mazur, submitted to Physical Review {\bf D}, February 1996.
\item{16}   I. Novikov, Ya. Zel'dovich, Pisma v ZETF, 1966; JETP Letters, 1966.
\item{17}   P. O. Mazur, Lectures given at XXXIV Cracow School of Theoretical 
Physics, June 1994, Zakopane, Poland; Lectures given at the Nordic School 
on High Energy Physics, January 1994, Spatind, Norway. 
\item{18}   A. Z. G\'orski, P. O. Mazur, and J. Szmigielski, submitted to 
Physics Letters {\bf B}, February 1996.

\end